# Spoof surface plasmon Fabry-Perot open resonators in a surface-wave photonic crystal

Zhen Gao[1], Fei Gao[1], Hongyi Xu[1], Youming Zhang[1], Baile Zhang*[1,2]

[1]*Division of Physics and Applied Physics, School of Physical and Mathematical Sciences, Nanyang Technological University, Singapore 637371, Singapore*

[2]*Centre for Disruptive Photonic Technologies, Nanyang Technological University, Singapore 637371, Singapore*

*\*Authors to whom correspondence should be addressed; E-mail: blzhang@ntu.edu.sg (B. Zhang)*

## Abstract

We report on the proposal and experimental realization of a spoof surface plasmon Fabry-Perot (FP) open resonator in a surface-wave photonic crystal. This surface-wave FP open resonator is formed by introducing a finite line defect in a surface-wave photonic crystal. The resonance frequencies of the surface-wave FP open resonator lie exactly within the forbidden band gap of the surface-wave photonic crystal and the FP open resonator uses this complete forbidden band gap to concentrate surface waves within a subwavelength cavity. Due to the complete forbidden band gap of the surface-wave photonic crystal, a new FP plasmonic resonance mode that exhibits monopolar features which is missing in traditional FP resonators and plasmonic resonators is demonstrated. Near-field response spectra and mode profiles are presented in the microwave regime to characterize properties of the proposed FP open resonator for spoof surface plasmons.

Surface plasmons (SPs), or surface electromagnetic waves that strongly couple to free electron oscillations at a metal-dielectric interface [1], are deemed promising candidates for the transport and manipulation of photons in subwavelength scales [2-4] owning to their tightly localized electromagnetic fields. In general, SPs exist either as propagating surface plasmon polaritons (SPPs) on an extended metal surface [1] or as localized surface plasmon polaritons (LSPs) on a finite metal surface, i.e., plasmonic particles [4]. In particular, such electron-photon coupled resonances, well known as localized surface plasmon resonances (LSPRs) on plasmonic particles, have resulted in unique and promising applications in photonic and biological technologies [4].

Generally, plasmonic particles with finite metal surfaces can be classified into two types: (1) plasmonic particles with size much smaller than the light wavelength (5-50 nm), where the optical behavior is mainly determined by resonances of dipole character; and (2) elongated plasmonic particles with size comparable to the wavelength (termed as "surface plasmon Fabry-Perot resonator"), which exhibit multipolar FP plasmonic resonance in the longitudinal direction [5-7]. Recently, to export the properties of LSPs to low frequencies (far-infrared, terahertz and microwave frequencies), the concept of spoof localized surface plasmons (spoof-LSPs) was proposed by periodically texturing plasmonic particle with closed surfaces much smaller than the wavelength [8] and experimentally verified on an ultrathin metallic disk at microwave frequencies [9-13]. However, up to now experimental investigations of multipolar spoof-LSPs supported on a surface plasmon FP open resonator have not been reported yet due to the serious radiation loss at the end-faces of the FP resonator with open surfaces.

More recently, by merging the waveguiding capability offered by photonic crystal [14] and the deep subwavelength nature of surface plasmons [15-19], the concept of surface-wave photonic crystal [20-25] was developed with superior properties including multi-directional splitting of surface waves with full bandwidth isolation [23], broadband high transmission through sharp bends [24], and slow wave devices [25].

Here, we realize a spoof surface plasmon Fabry-Perot open resonator by introducing a finite line defect in a surface-wave photonic crystal. In contrast to conventional spoof-LSPs supported on textured closed metal surfaces which originate from the interference of clockwise and counterclockwise propagating surface modes [8-12], the FP open resonator modes, i.e., standing surface waves tightly localized in the finite line defect, originate from the end-face reflections due to the complete forbidden band gap of the surrounding surface-wave photonic crystal for surface modes.

Moreover, unlike conventional FP resonators [5-7] and plasmonic resonators [8-13] standing in a homogeneous medium (e.g., air) where waves can propagate, both the field confinement and the quality factors of the spoof surface plasmon FP open resonators are enhanced due to the forbidden band gap offered by the surrounding surface-wave photonic crystal. A new FP resonance mode with monopolar features, which is missed in previous FP resonators [5-7], is emerged due to the forbidden band gap of the surrounding surface-wave photonic crystal. We also demonstrate that the main characteristics of the FP open resonators, the total numbers and the resonance frequencies of the FP resonance modes, can be conveniently engineered through changing the defect numbers (or total length) of the open resonator, rather than changing its whole geometry in a re-fabrication.

To experimentally demonstrate the existence of the surface resonant states of the FP open resonator, we fabricate a surface-wave photonic crystal with a square array (13 × 13) of cylindrical aluminum pillars standing on a flat metal surface, as shown in Fig. 1, in which period $d$ is 5.0 mm, and the height $H$ and radius $r$ of the cylindrical pillars are 5.0 mm and 1.25 mm, respectively. Such a surface-wave photonic crystal can exhibit complete forbidden band gap for surface waves [22-26]. Previous Finite Integration Technique (FIT) eigenmode simulation has revealed a complete forbidden band gap from 12.6 GHz to 27 GHz for the current surface-wave photonic crystal [22].

To realize a spoof surface plasmon FP open resonator, a finite line-defect can be constructed by shortening the height of several cylindrical pillars from $H = 5.0$ mm to $h = 4.3$ mm in the middle of the surface-wave photonic crystal, thus the resonance frequencies of the FP open resonator are located within the forbidden band gap of the surrounding surface-wave photonic crystal. We first construct a finite line defect with seven shortened pillars (N = 7) in the middle of the surface-wave photonic crystal. The experimental setup is shown in the inset of Fig. 2(a), a monopole antenna as the source near the entrance surface to excite the FP open resonator and another monopole antenna as the probe near the end surface of the finite line-defect to measure its near-field response spectrum (S-parameter S21). The measured near-field response spectrum is shown in Fig. 2(b), from which seven resonance peaks marked as $M_0$-$M_6$ can be clearly observed within the photonic band gap of the surrounding surface-wave photonic crystal (the purple region). For completeness, we also measured the near-field response spectrum (grey line in Fig. 2(b)) of the domino-plasmon open resonator by removing the surrounding surface-wave photonic crystal; the resonances are rather weak due to the scattering at the end-faces of the domino-plasmon open resonator.

These FP resonance modes are standing surface waves along the finite line-defect which exist whenever an integer of half the surface plasmon wavelength equals to the length of finite line-defect. Thus we can write the resonance condition of the FP open resonator as: $K_{spp} \times l = m \times \pi - \emptyset$, where $K_{spp}$ is the wavevector of the surface waves, $l = N \times d$ is the length of the FP open resonator (N is the number of the defect pillars), $\emptyset$ is the reflection phase of the entrance (or end) face of the FP open resonator, and m is an integer denoting the order of the resonance modes. Note that the end-face reflection phase shift at both terminations ($\emptyset$) can be approximately experimentally retrieved from the measured resonance frequencies and the dispersion relation of the line-defect waveguide [24]. For instance, for the first-order resonance mode (m = 1) of the FP open resonator with defect pillar number N = 7, its measured resonance frequency is 12.86 GHz, then we can get the wavevecor ($K_{spp}$) from the dispersion relation that $K_{spp} = 0.4\,\pi/d$, and the total length of the resonator $l = N \times d$, where d is the periodicity of the line defect. Thus we can deduce that the reflection phase for the first-order resonance mode of the FP open resonator with defect pillar number N = 7 is about $\emptyset = 0.2\pi$.

Direct imaging of the surface plasmon near-fields on the FP open resonator are performed using a microwave near-field scanning system [27-29] to directly verify this resonance condition. Fig. 2(c) shows the measured $E_z$ field distributions of the seven resonance modes in a transvers plane 1 mm above the top surface of the FP open resonator. As expected, the standing surface waves occur at each resonance peak in the near-field response spectrum of Fig. 2(b) due to the multiple spoof surface plasmon reflections at the entrance and end faces of the FP open resonator. One can easily see that the order (m) of the seven resonance modes are 0, 1, 2, 3, 4, 5, and 6 respectively. Interestingly, we observe a zeroth-order mode ($M_0$) for this spoof surface plasmon FP open resonator, which is missed in all previous conventional FP resonators [5-7]. We attribute this new emerged zeroth-order FP mode to the tight confinement provided by the complete forbidden band gap of the surrounding surface-wave photonic crystal.

We further decrease the defect pillar numbers of the FP open resonator to N = 5 [inset of Fig. 3(a)]. For shorter finite line-defect length, we expect to have FP resonances that are spectrally farther apart in analogy to the larger mode spacing in a conventional shorter FP resonator. We use the same experimental setup (two monopole antennas as the source and probe are indicated with black arrows) to measure its near-field response spectrum, as shown in Fig. 3(a), from which five resonance peaks with larger mode spacing can be observed. The corresponding measured near-field distributions ($E_z$) of the five resonance peaks ($M_0$-$M_4$) are presented in Fig. 3(b). It is clear that the mode order is 0, 1, 2, 3, and 4 for the five resonance modes.

Compared with the longer FP open resonator in Fig. 2, several conclusions can be drawn from the measured near-field response spectrum and mapped near-field mode profiles. First, the total numbers of FP resonance modes are decreased from seven to five, and two highest order FP modes ($M_5$, $M_6$) disappear from the near-field response spectrum when the defect pillar numbers decrease from N = 7 to N = 5. That is because, in order to maintain the same mode order ($m$), decreasing defect numbers (N) must be accompanied with increasing wavevector ($K_{spp}$) of propagating spoof surface plasmon. Thus for the FP open resonator with defect numbers N = 5, when the mode order is higher than five ($m \geq 5$), the wavevector ($K_{spp}$) will beyond the edge of the first Brillion zone of the spoof surface plasmon dispersion [10] in an infinite line-defect waveguide, thus the propagation of spoof surface plasmon will be cutoff and the FP resonances disappear. Second, for the same order mode (for example, m = 0), when the defect numbers decrease from N = 7 to N = 5, the resonance frequencies of the zeroth-order mode ($M_0$) increase from 12.65 GHz to 12.75 GHz. This is because, decreasing the defect numbers (N) shrinks the total length ($l$) of the FP open resonator, and the wavevector of spoof surface plasmon ($K_{spp}$) needs to be increased to meet the resonance condition of the fixed mode order (m = 0). Thus the resonance frequencies of the FP resonator modes with the fixed mode order also increase, following the dispersion relation of spoof surface plasmons propagating along the line-defect waveguide in a surface-wave photonic crystal [24]. These two conclusions indicate the tunability of these spoof surface plasmon FP open resonators by simply changing their defect numbers.

When we continue to decrease the defect pillar numbers to N = 3, as shown in the inset of Fig. 4(a), we can only observe three resonance peaks in the measured near-field response spectrum. The measured near-field mode profiles correspond to the three resonance peaks ($M_0$-$M_2$) reveal that their mode orders are 0, 1 and 2, respectively. For completeness we further decrease the defect number to N = 1, which is a point defect [22,26] in surface-wave photonic crystal, as shown in the inset of Fig. 5(a). It is obvious that we can only observe a single resonance peak in the measured near-filed response spectrum. This single resonance peak corresponds to the zeroth-order resonance mode ($M_0$), whose measured mode profile ($E_z$) is shown in Fig. 5(b), matched well with previous reported results [22,26].

Note that the zeroth-order mode ($M_0$) always exists no matter how the number of the defect pillars changes because of the typical point defect cavity properties of photonic crystal [14]. However, for the traditional plasmonic FP resonators [5-7], if we continue to decrease the resonator length, they will gradually reduce to small plasmonic particles with size much smaller to the light wavelength, and their resonance properties are mainly characterized by dipolar features. The emergence of the zeroth-order FP plasmonic mode ($M_0$) mainly comes from the advantage of combination of photonic crystal and subwavelength nature of surface plasmons.

In conclusion, we have proposed and experimentally demonstrated a finite line-defect in surface-wave photonic crystal works as a new configuration of tunable spoof surface plasmon FP open resonators in the microwave frequencies. We also show that both the total numbers and resonance frequencies of the FP resonance modes can be conveniently tuned by changing the defect pillar numbers of the open resonator, rather than its complete geometry, simplifying the design of FP open resonators to only one degree of freedom. Moreover, we observe a zeroth-order mode for this spoof surface plasmon FP open resonator, which is missed in all previous traditional FP resonators and plasmonic resoantors, due to the forbidden band gap provided by the surface-wave photonic crystal. These results will be useful for future plasmonic resonator design that are important for integrated plasmonic circuits.

## Acknowledgements

This work was sponsored by the NTU Start-Up Grants, Singapore Ministry of Education under Grant No. MOE2015-T2-1-070 and MOE2011-T3-1-005.

**References**

1. W. L. Barnes, A. Dereux, and T. W. Ebbesen, Nature **424**, 824 (2003).
2. S. I. Bozhevolnyi, V. S. Volkov, E. Devaux, J. Y. Laluet, and T. W. Ebbesen, Nature **440**, 508-511 (2006).
3. E. Moreno, S. G. Rodrigo, S. I. Bozhevolnyi, L. Martín-Moreno, and F. J. García-Vidal, Phys. Rev. Lett. **100**, 023901 (2008).
4. U. Kreibig and M. Vollmer, *Optical Properties of Metal Clusters* (Springer, Berlin, 1995).
5. H. Ditlbacher, A. Hohenau, D. Wagner, U. Kreibig, M. Rogers, F. Hofer, F. R. Aussenegg, and J. R. Krenn, Phys. Rev. Lett. **95**, 257403 (2005)
6. E. Vesseur, R. Waele, M. Kuttge, and A. Polman, Nano Lett., **7** (9), 2843–2846, (2007).
7. P. Ghenuche, S. Cherukulappurath, T. H. Taminiau, N. F. van Hulst, and R. Quidant, Phys. Rev. Lett. **101**, 116805 (2008).
8. A. Pors, E. Moreno, L. Martin-Moreno, J. B. Pendry, and F. J. Garcia-Vidal, Phys. Rev. Lett. **108**, 223905 (2012).
9. X. P. Shen and T. J. Cui, Laser Photonics Rev. **8**, 137-145 (2014).
10. F. Gao, Z. Gao, X. H. Shi, Z. J. Yang, X. Lin, and B. L. Zhang, Opt. Express **23**, 6896-6902 (2015).
11. F. Gao, Z. Gao, Y. M. Zhang, X. H. Shi, Z. J. Yang and B. L. Zhang, Laser Photonics Rev. **9**, 571-576 (2015).
12. Z. Gao, F. Gao, K. K. Shastri, and B. L. Zhang, Sci. Rep. **6**, 25576 (2016).
13. Z. Gao, F. Gao, H. Y. Xu, Y. M. Zhang, B. L. Zhang, Opt. Lett. **41**, 002181 (2016).
14. J. D. Joannopoulos, R. D. Meade, and J. N. Winn, *Photonic Crystals: Molding the Flow of Light. Princeton*. (Princeton University Press, 2008.)
15. J. B. Pendry, A. J. Holden, W. J. Stewart, and I. Youngs, Phys. Rev. Lett. **76**, 4773 (1996).
16. J. B. Pendry, L. Martín-Moreno, and F. J. Garcia-Vidal, Science **305**, 847-848 (2004).
17. A. P. Hibbins, B. R. Evans, and J. R. Sambles, Science **308**, 670-672 (2005).
18. D. Martin-Cano, M. L. Nesterov, A. I. Fernandez-Dominguez, F. J. Garcia-Vidal, L. Martin-Moreno, and E. Moreno, Opt. Express **18**, 754-764 (2010).

19. X. P. Shen, T. J. Cui, D. Martin-Cano, and F. J. Garcia-Vidal, Proc. Natl. Acad. Sci. **110**, 40-45 (2013).
20. F. Lemoult, N. Kaina, M. Fink, and G. Lerosey, Nat. Phys. **9**, 55-60 (2013).
21. N. Kaina, F. Lemoult, M. Fink, and G. Lerosey, Appl. Phys. Lett. **102**, 144104 (2013).
22. Z. Gao, F. Gao, and B. L. Zhang, Appl. Phys. Lett. **108**, 041105 (2016).
23. Z. Gao, F. Gao, and B. L. Zhang, Appl. Phys. Lett. **108**, 111107 (2016).
24. Z. Gao and B. L. Zhang, arXiv:1604.06872 (2016).
25. N. Kaina, A. Causier, Y. Bourlier, M. Fink, T. Berthelot, G. Lerosey, arXiv:1604.08117 (2016).
26. S. H. Kim, S. S. Oh, K. J. Kim, J. E. Kim, H. Y. Park, O. Hess, and C. S. Kee, Phys. Rev. B **91**, 035116 (2015).
27. Z. Gao, F. Gao, Y. M. Zhang, B. L. Zhang, Phys. Rev **B**. 93, 195410 (2016).
28. Z. Gao, F. Gao, Y. M. Zhang, X. H. Shi , Z. J. Yang , B. L. Zhang, Appl. Phys. Lett. **107**, 041118 (2015).
29. Z. Gao, F. Gao, Y. M. Zhang, B. L. Zhang, Appl. Phys. Lett. **107**, 191103 (2016).

**Figures and captions**

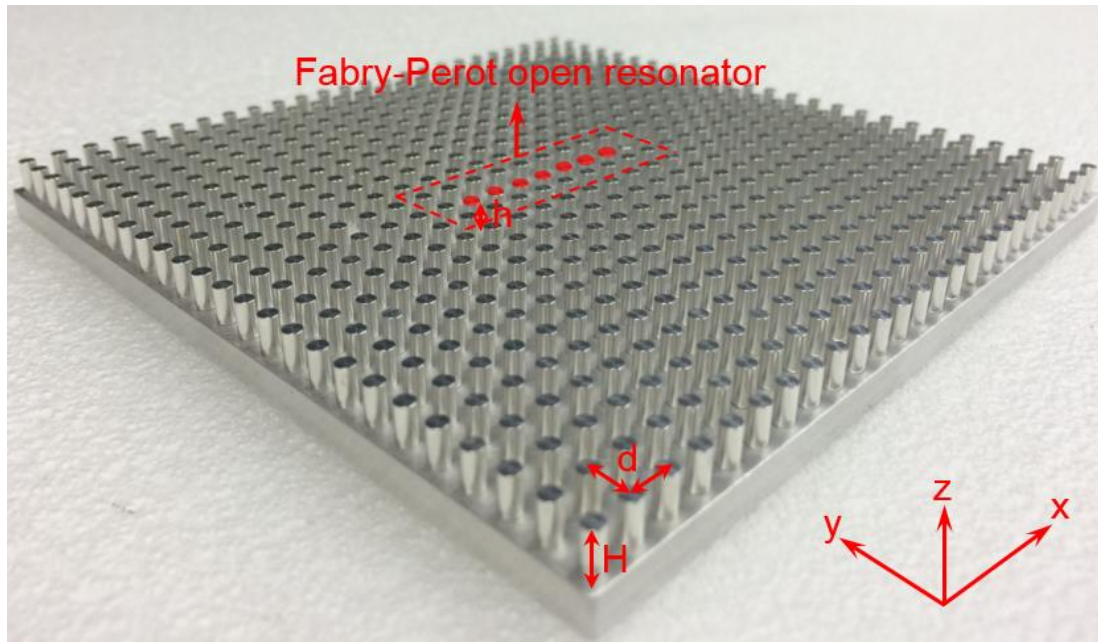

**FIG. 1.** (a) Photography of the Fabry-Perot open resonators (dashed rectangle) in a surface-wave photonic crystal that consists of a square array of cylindrical aluminum pillars with radius $r = 1.25$ mm, height $H = 5.0$ mm, and periodicity $d = 5.0$ mm, all standing on a metal surface. The spoof surface plasmon Fabry-Perot open resonators are created by shortening several cylindrical pillars (white dots) from height $H = 5.0$ mm to $h = 4.3$ mm in the middle of the surface-wave photonic crystal.

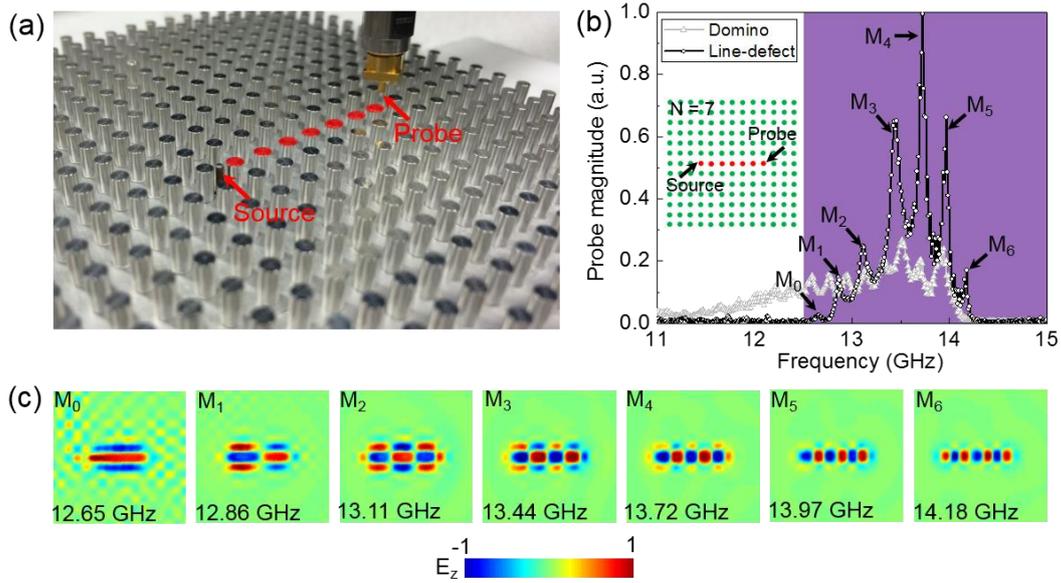

**FIG. 2.** (a) Experiment setup to measure the near-field response spectrum and near-field mode profiles of the Fabry-Perot open resonator which consists of seven defect pillars (white dots). Two monopole antennas as the source and probe are indicated by two black arrows. (b) Measured near-field response spectrum of the FP open resonator (black line). The near-field response spectrum of a finite domino waveguide (grey line) is also measured for comparison. Inset shows the schematic of the open resonator. Purple region indicates the forbidden band gap of the surrounding surface-wave photonic crystal. (c) Observed field patterns ($E_z$) of the seven corresponding resonance peaks ($M_0$-$M_6$) in a transverse plane 1 mm above the top surface of the FP resonator.

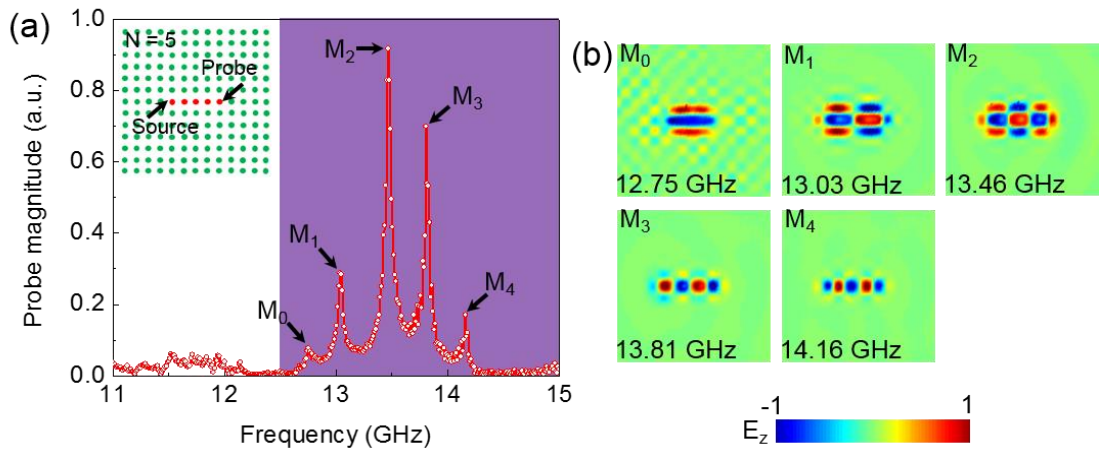

**FIG. 3.** (a) Measured near-field response spectrum of a Fabry-Perot open resonator consisting of five defect pillars. Inset shows the schematic of open resonator. (b) Measured field patterns ($E_z$) of the five resonance modes ($M_0$-$M_4$) in a transverse plane 1 mm above the top surface of the FP resonator.

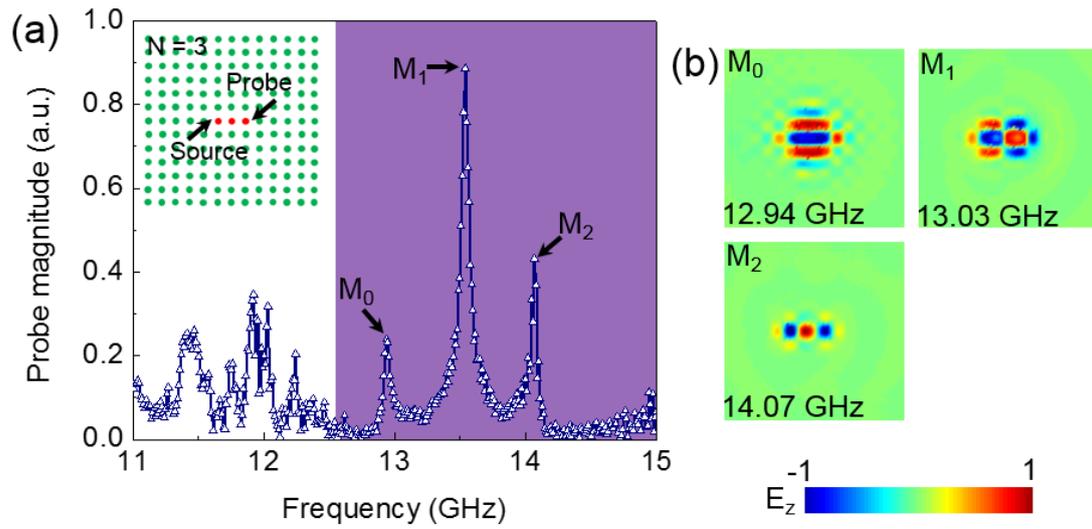

**FIG. 4.** (a) Measured near-field response spectrum of a Fabry-Perot open resonator consisting of three defect pillars. Inset shows the schematic of open resonator. (b) Measured field patterns ($E_z$) of the three resonance modes ($M_0$-$M_2$) in a transverse plane 1 mm above the top surface of the FP resonator.

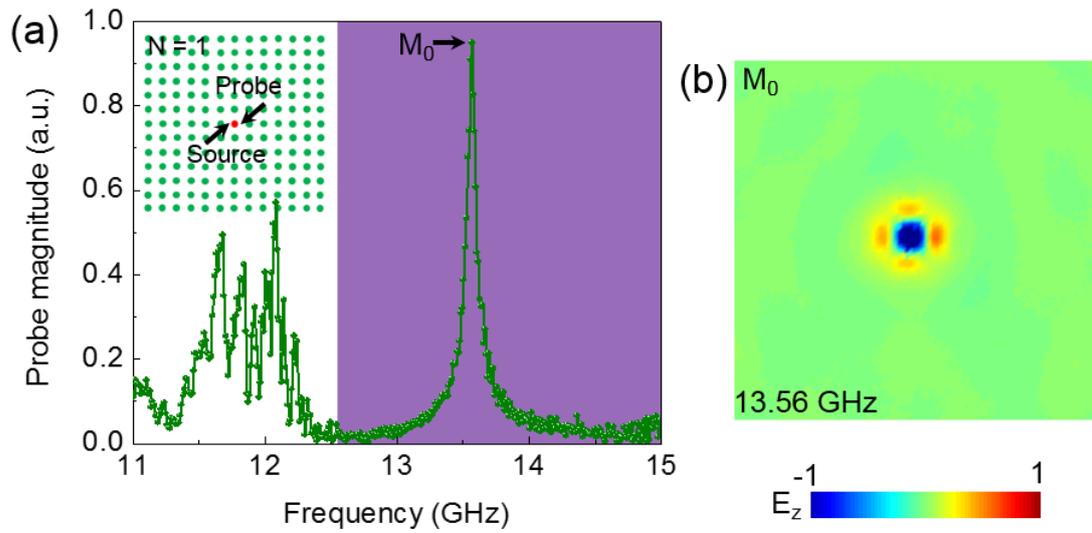

**FIG. 5.** (a) Measured near-field response spectrum of a single point defect in a surface-wave photonic crystal. Inset shows the schematic of the point defect. (b) Measured field pattern ($E_z$) of the single resonance modes ($M_0$) in a transverse plane 1 mm above the top surface of the point defect.